\documentstyle[twocolumn,eqsecnum,aps]{revtex}
\begin{document}
\draft
\preprint{KYUSHU-HET-41}
\title{Theta Dependence of Meson Masses in the Small Mass Limit \\
of the Massive Schwinger Model}
\author{Matthias Burkardt}
\address{Department of Physics\\
New Mexico State University\\
Las Cruces, NM 88003-0001\\U.S.A.}
\author{Koji Harada}
\address{Department of Physics\\
Kyushu University\\
Fukuoka 812-81\\
Japan}
%\date{\today}
\maketitle
\begin{abstract}
We present a continuum formulation for $\theta$-vacua in the massive
Schwinger model on the light-front, where $\theta$ enters as a background 
electric field. The effective coupling of the external field is partially 
screened due to vacuum polarization processes.
For small fermion masses and small $\theta$ we calculate
the mass of the meson
and find agreement with results from
bosonization.
\end{abstract}
%\pacs{Valid PACS appear here}
%{\tt$\backslash$\string pacs\{\}} should always be input,
%even if empty.}
\narrowtext

\section{Introduction}
Light-Front (LF) coordinates are natural coordinates for
studying many high energy scattering processes, since such processes
are often dominated by correlations along light-like directions
\cite{mb:adv,dave:stan}. For recent reviews on this subject, see
for example Refs. \cite{mb:adv,big:guy}

The LF framework is very useful also for studying low-energy,
nonperturbative physics such as bound states because 
field theories quantized on the light-front have the vacuum 
which appears to be trivial (at least
as long as zero-mode degrees of freedom are neglected).
But how can the LF trivial vacuum incorporate with 
non-trivial structures (e.g., spontaneous symmetry breaking) 
of the vacuum? 
Presumably all the vacuum structures would be extracted from the
zero-mode dynamics, if we could correctly take it into account.
It is unfortunately extremely difficult to do it. On the other
hand, there exists now plenty of evidence that there is no conflict
between trivial vacua on the LF and nontrivial vacua in an equal
time formulation \cite{mb:nuss}, {\it provided} one works and interprets
the LF Hamiltonian as an effective Hamiltonian in which zero-mode 
degrees of freedom have been integrated out (as opposed to just 
ommitted). 
The point is that it seems that effects of the zero mode dynamics
can really be simulated by a set of ``counterterms,'' though their
precise forms and the strengths are not easily determined.

Nevertheless, it is still common lore that such an
effective LF Hamiltonian approach is not sufficient to account for
topologically nontrivial effects. In particular it is generally 
assumed that $\theta$ vacua in the LF framework can only be described
if the zero-modes of the gauge field are included as dynamical
degrees of freedom \cite{gary,lfnulls}.

In order to investigate this issue, we are investigating the massive
Schwinger model which is known to have $\theta$ vacua \cite{cjs}
and where the properties of mesons are dependant on the value of 
$\theta$ \cite{coleman}. In the equal time formulation of the model,
there are (at least) two complementary approaches to describe the 
physics of $\theta$ vacua: on the one hand, one can formulate the
model on a finte interval with periodic boundary conditions and
introduce dynamical zero-mode degrees of freedom for the gauge field
\cite{manton}. In this approach, the parameter $\theta$ appears in
a way very similar to the lattice momentum in Bloch waves familiar
from solid state physics. On the other hand, one can interpret
$\theta$ as an external field, generated by ``intergalactic capacitor
plates'' at infinity \cite{coleman}. In the latter approach, where the
1+1 dimensional world is not a circle but the infinite line, $\theta$
is not a dynamical degree of freedom, but instead appears as an
integration constant when solving Poisson's equation.

LF formulations of $\theta$-vacua in the Schwinger model which
put the system in a box with periodic boundary conditions and where 
the $\theta$ vacuum is constructed as a Bloch state can be found in
Refs. \cite{lfnulls,hkw} ($m=0$) and \cite{koji} ($m\neq 0$). 
% koji
(See also Ref. \cite{lubo}.)
One finds that
$\theta$ vacua can be understood as zero-mode dynamics and $\theta$ plays
the role of a Bloch momentum. In particular, the periodicity of the
physics in $\theta$ can be quite easily understood, though it is due to
non-trivial dynamics (pair creation). Even though the vacuum and meson
equations have been obtained, spectra and wave functions have not been
calculated explicitly because the continuum limit of the equations is
singular (or ambigous). It is therefore desirable to have an alternate
formulation which allows to perform explicit calculations.

The approach to $\theta$ vacua in the LF formulation which we will
pursue in this paper is complementary to the ``Bloch wave approach''
described above and is in fact very similar to Coleman's formulation of the
problem in an equal time framework: we introduce the $\theta$ 
parameter as an external field, generated by external charges at
infinity and then we study the behavior of mesons under the influence
of such an external field.
The paper is organized as follows: first we derive the effective
interaction term caused by such an external electric field.
In the rest of the paper we investigate the influence of this field
on mesons by approximating mesons as fermion -- anti-fermion pairs.
In particular, we study the boundary behavior of meson wave-functions,
screening effects and the chiral limit of the model.

\section{$\theta$ vacua as effective background fields}
Let us start with the Lagrangian of the massive Schwinger model
\begin{equation}
{\cal L}=-{1\over 4}F_{\mu\nu}F^{\mu\nu}
+\bar\psi[\gamma^\mu(i\partial_\mu-eA_\mu)-m]\psi.
\end{equation}
We quantize the model on the light-cone, i.e., by regarding $x^+
=(x^0+x^1)/\sqrt{2}$ as the ``time.'' In the $A^+=0$ gauge\footnote{
We neglect explicit zero-mode degrees of freedom.},
equations of motion can be derived in the usual way. In particular,
we find that $A^-$ satisfies the following Poisson equation,
\begin{equation}
-\partial_-^2 A^- = ej^+,
\label{poisson}
\end{equation}
where $j^+=\sqrt{2}:\psi^\dagger_R\psi_R:$, $\psi=(\psi_R, \psi_L)^T$.
For the notation used in this paper, see Ref.\cite{6bhot}.

Following Coleman, we introduce the $\theta$ parameter into the
massive Schwinger model as an external background field by
including an appropriate integration constant in the solution to 
Eq.(\ref{poisson}),
\begin{equation}
A^-(x^-) =  -\frac{e}{2} \int_{-\infty}^\infty \,\,dy^- |x^--y^-|j^+(y^-)
\, - \,\frac{e\theta}{2\pi} x^- ,
\end{equation}
yielding an additional interaction term in the LF Hamiltonian\footnote{
We have used translational invariance, i.e.,
\begin{eqnarray}
\int dx^-{1\over \partial_-}j^+
&=&{1\over 2}\int dx^-dy^-\epsilon(x^--y^-)j^+(y^-) \nonumber \\
&=& {1\over 2}\int dz^-\epsilon(z^-)Q=0 \nonumber
\end{eqnarray}
when $Q\equiv \int dx^- j^+=0$.
}
\begin{equation}
\delta P^- = -\frac{e^2\theta}{2\pi} \int_{-\infty}^\infty dx^- j^+(x^-)x^- .
\end{equation}

In momentum space, this additional term appears as the derivative of
the current operator at zero momentum, i.e.,
\begin{equation}
\delta P^- = -\frac{ie^2\theta}{2\pi}\left. \frac{d}{dq^+} j^+(q^+)
\right|_{q^+=0},
\label{eq:p-}
\end{equation}
where for $q^+>0$
\begin{eqnarray}
j^+(q^+) &\equiv& 
\int dx^- e^{-iq^+x^-}j^+(x^-)\nonumber\\
&=&\int_0^\infty \!\!\!{dk^+\over 2\pi\sqrt{k^+(k^++q^+)}}
 \left[ b^\dagger (k^++q^+)b(k^+)\right. \nonumber \\
& &\left.\qquad\quad {}-d^\dagger (k^++q^+)d(k^+) \right] \nonumber\\
& &+\int_0^{q^+} \!\!\!{dk^+\over 2\pi\sqrt{k^+(q^+-k^+)}}
b^\dagger(k^+) d^\dagger(q^+-k^+)
\nonumber\\
&=& j^+_{diag}+j^+_{pair}
\label{eq:j+}
\end{eqnarray}
and analogously for $q^+<0$. The $b$ and $d$ are the usual 
destruction operators for fermions and anti-fermions respectively and,
by definition, $j^+_{diag}$ ($j^+_{pair}$) are those terms in the 
current operator which are diagonal (off-diagonal) in Fock space.
The contributions of $j^+_{diag}$ to $\delta P^-$ [Eq. (\ref{eq:p-})] are 
straightforward to evaluate, yielding
\begin{eqnarray}
\delta P^-_{diag} &=& {ie^2\theta \over 2\pi}
\!\int_0^\infty \!\!\!{dk^+ dp^+\over \sqrt{k^+p^+}}
\delta'(k^+\!\!-\!\!p^+) \nonumber \\
& &\qquad\times[b^\dagger(k^+)b(p^+)-d^\dagger(k^+)d(p^+)],
\label{eq:p-naive}
\end{eqnarray}
i.e., a derivative coupling with opposite signs for fermions and
anti-fermions.

How can $j^+_{pair}$ in $\delta P^-$ affect the meson state? 
First observe that the (anti-)fermion must have non-zero momentum $q^+$
in order for  $j^+_{pair}(q^+)$ to contribute. Because wave functions
in the massive Schwinger model vanish when one of the momenta goes to zero,
one might think that $j^+_{pair}(q^+)$ with $q^+\rightarrow 0$ (hence
$\delta P^-_{pair}$) is unimportant.
However, we can show that they do not vanish when the momenta of a fermion
and an anti-fermion go to zero {\it simultaneously}. The details of the
argument will be presented in Ref. \cite{us} and in this letter we will
restrict ourselves to analyzing the consequences for the coupling
of a meson to a constant background field. 

\begin{figure}
%\begin{Large}
\unitlength1.cm
\begin{picture}(15,7.)(.5,-8.5)
\includegraphics{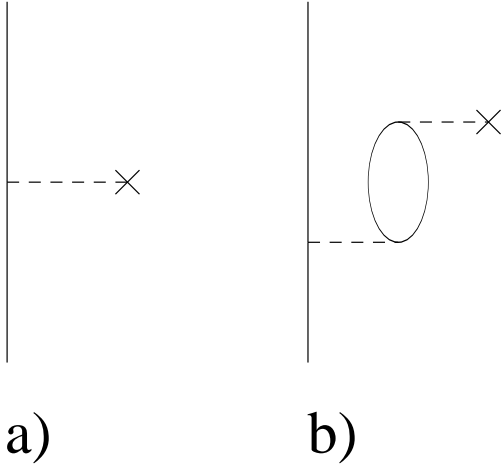}
\end{picture}
%\end{Large}
\caption{Diagrams contributing to the interaction of a fermion to an 
external charge.
(a) tree level coupling, (b) vacuum polarization
diagram where the pair creation piece
in the current operator contributes.
}
\label{fig:j+}
\end{figure}

In order to investigate whether this has a nontrivial effect on
the matrix elements of $\left.\frac{d}{dq^+}j^+\right|_{q^+=0}$, 
let us consider
the matrix elements of $j^+$ for small but nonzero momentum transfer.
As an example, compare the two diagrams in Fig.~\ref{fig:j+}, 
which contribute to the coupling of a fermion to an external charge. 
Fig. \ref{fig:j+}a is just the bare tree level coupling, yielding
\begin{equation}
V_a = \frac{1}{{q^+}^2},
\label{eq:va}
\end{equation}
while the one loop diagram [Fig. \ref{fig:j+}b] yields
\begin{equation}
V_b = \frac{e^2}{2\pi}\frac{1}{{q^+}^4}
\int_0^{q^+} \frac{dk^+}{q^- - \frac{m^2}{2k^+} - \frac{m^2}{2(q^+-k^+)}}
\stackrel{q^+\rightarrow 0}{\longrightarrow}
- \frac{e^2}{\pi}\frac{1}{6m^2}\frac{1}{{q^+}^2}
\label{eq:vb}
\end{equation}
(since we are interested in static external sources, we may neglect
$q^-$ in $V_b$). Clearly, Eq. (\ref{eq:vb}) is of the same order
in $q^+$ as Eq. (\ref{eq:va}) and may thus not be neglected
--- even in the limit $q^+ \rightarrow 0$. Thus, the bad news is that
even for small $q^+$, pair creation terms cannot be neglected in the
matrix elements of the current operator. However, the good news is 
that for $q^+\rightarrow 0$ (which is the case we are interested in for
the coupling to an external constant electric field)
the relevant contribution can be calculated from vacuum polarization
diagrams only\footnote{We were not able to give a rigorous proof of
this result, but convinced ourselves about the correctness by studying
many different perturbative diagrams.}. 
The nonperturbative contribution from all vacuum polarization diagrams can 
be summed
up by inserting a complete set of meson states, yielding
\begin{equation}
V_{vac-pol}=-\frac{1}{{q^+}^2}
\frac{e^2}{\pi} \sum_n \frac{g_V(n)^2}{\mu_n^2},
\end{equation}
where 
$ g_V(n) = \int_0^1 dx \psi_n(x) $
and $\psi_n(x)$ is the wave function of the n-th meson with mass
$\mu_n$ in the two particle sector.
Thus while it is incorrect to leave out couplings of the external 
current which are off-diagonal in Fock space\footnote{Note 
that since it is the anomaly which is responsible for the nonvanishing
contribution of the off-diagonal pieces, it is only the $U(1)$ current
which is affected by this result in a multi-flavor version of the model.},
we can take them into account effectively by renormalizing the 
all couplings to the external field according to the rule
\begin{equation}
\theta \longrightarrow \theta_{eff} \equiv \theta
\left[1-\frac{e^2}{\pi}\sum_n \frac{g_V(n)^2}{\mu_n^2}\right] .
\label{eq:thetaeff}
\end{equation}
At the same time, all matrix elements that are off-diagonal in
Fock-space may be omitted (this is what makes the effective interaction
effective), i.e. we find the amazingly simple result that
the naive interaction term (\ref{eq:p-naive}) becomes correct provided one 
replaces $\theta$ by $\theta_{eff}$.

Note that the diagrams with a ``chain'' of vacuum polarizations ending 
at the external source are special in the sense that such diagrams are 
not generated by the usual perturbation theory with $\delta P^-$ as
the perturbation. It is the interplay of $\delta P^-$ and the
(so-called ``Fork'') interactions already in the Hamiltonian that
generates these diagrams. It is however still true that the whole set
of these diagrams vanishes when $\delta P^-=0$.

As a side remark, we should emphasize that in principle
the screening factor $\theta_{eff} /\theta$ itself depends on 
$\theta_{eff}$ since it
is $\theta_{eff}$ which enters the LF-Hamiltonian and thus 
determines the masses and coupling constants of mesons that appear
in the sum in Eq. (\ref{eq:thetaeff}). However, we will neglect this
effect here and evaluate the screening factor at $\theta=0$, which
will effectively limit the validity of our results to small values of 
$\theta$.

\section{The $\theta$ parameter in the two particle sector}
In order to illustrate the physics of the $\theta$ parameter, let us now
focus our attention to the 2-particle sector of the massive Schwinger
model, which is known to provide an excellent approximation for the
lightest meson \cite{berg}.
After projecting the effective LF Hamiltonian in the presence of
an external background field onto the two particle sector, one finds
the following equation of motion for mesons
\begin{eqnarray}
\mu^2 \psi (x) &=& \frac{m^2-1}{x(1-x)}\psi (x)
+ \int_0^1 dy \frac{1}{x-y} \frac{d}{dy}\psi (y) 
\nonumber\\
&+& \int_0^1 dy \psi (y) + i\theta_{eff} \frac{d}{dx}\psi (x),
\label{eq:bh}
\end{eqnarray}
where a principal value prescription for the singular integral is
implied. Here and in the following we will use units where $e^2/\pi=1$.

The most dramatic modification compared to the Bergknoff Hamiltonian
($\theta=0$) is that above Hamiltonian is complex (but still Hermitian).
The eigenfunctions satisfy
\begin{equation}
\psi(1-x) = \pm \psi^*(x),
\end{equation}
which results from charge conjugation invariance. At the boundary $x=0,1$
a self-consistent ansatz shows that
$\psi(x)$ vaishes like $x^\beta$ and $(1-x)^{\beta^*}$ respectively, where 
$\beta$ is the solution to
\begin{equation}
m^2-1 + \pi \beta \cot (\pi \beta) + i \beta \theta_{eff} =0
\label{eq:bbb}
\end{equation}
with $\Re \beta \in (0;1)$. Obviously Eq. (\ref{eq:bbb}) can be satisfied 
if and only if one allows $\beta$
to be complex. For $\theta_{eff} =0$ one recovers 't Hooft's boundary
condition $\pi \beta \cot (\pi \beta)=1-m^2$ \cite{thooft},
which has only real solutions $\beta_n$ located at
$n<\beta_n<n+1$. For the end-point behavior only $\beta_0$ is important.
In the following we will briefly discuss how $\beta_n$ --- the solutions to Eq.
(\ref{eq:bbb}) change as a function of $\theta_{eff}$. Detailled proofs
can be found in Ref. \cite{us}. 

For fixed $m$, as one increases $\theta_{eff}$, the imaginary part of
$\beta$ increases until at a certain value 
$\theta_{eff}=\theta^{crit}_{eff}$,
$\beta$ becomes purely imaginary and
the solutions to the Bergknoff equation with $\theta$-term become tachyonic.
For $m^2\geq 1$, one finds $\theta_{eff}^{crit}=\pi$, but for $0<m^2<1$,
$\theta_{eff}^{crit}$ slowly decreases to zero. In the chiral limit
one can see this explicitly using (\ref{eq:bbb})
\begin{equation}
\frac{(\pi \beta)^2}{3} - i \beta \theta_{eff} - m^2 =0
\end{equation}
and thus
\begin{equation}
\frac{\pi}{\sqrt{3}} \beta = \sqrt{m^2-\frac{3}{4\pi^2}\theta_{eff}^2} 
+ i \frac{\sqrt{3}}{2\pi}\theta_{eff} ,
\label{eq:bchiral}  
\end{equation}
yielding
\begin{equation}
\theta_{eff}^{crit} = \frac{2\pi}{\sqrt{3}} m .
\end{equation}

\section{Variational calculation in the chiral limit}
For $m=0$, the screening factor (\ref{eq:thetaeff}) vanishes and thus
$\theta_{eff}=0$ regardless of the value of $\theta$. 
Since only
$\theta_{eff}$ enters the Hamiltonian for mesons, we thus confirm 
that meson masses become $\theta$-independent for zero fermion masses.

Much more interesting is the limit of small but nonvanishing fermion
masses. In this case numerical calculations of the meson spectrum
become very tricky because of the singular behavior of the meson wave 
functions at the boundary. Furthermore, especially for values of
$|\theta_{eff}| \approx \pi$, pair creation (which we are suppressing for
simplicity) is expected to become very important. 

Given that $\psi$ has to satisfy the boundary conditions
$ \psi (x) \rightarrow x^\beta$ and 
$(1-x)^{\beta^*}$ for $x\rightarrow 0$ and $1$,
where $\beta$ is determined from Eq. (\ref{eq:bbb}), it would be natural
to make a variational ansatz of the form
\begin{equation}
\psi (x) = x^\beta (1-x)^{\beta^*} .
\label{eq:psi1}
\end{equation}
However, we have not been able to derive analytic expressions for matrix
elements of the interaction with this ansatz for general (i.e. complex)
values of $\beta$. For $|\beta| \ll 1$, an ansatz which has the same
end-point behavior as Eq. (\ref{eq:bbb}), and which is also 
[like Eq. (\ref{eq:psi1})] nearly constant for intermediate values of $x$
is given by
\begin{equation}
\psi (x) = 
x^\beta (1-x)^{1-\beta}+x^{1-\beta^*} (1-x)^{\beta^*}
\label{eq:psi2} .
\end{equation}
The main advantage of this modified ansatz is that it not only satisfies
the right boundary conditions %(\ref{eq:bbbb}) 
but also leads to analytically calculable matrix elements,
using \cite{grabstein}
\begin{equation}
\int_0^1 dy \frac{y^\nu (1-y)^{-\nu}}{y-x}=
\frac{\pi}{\sin (\nu \pi)}
\left[ 1 - \cos (\nu \pi)x^\nu (1-x)^{-\nu}\right]
\label{eq:int}
\end{equation}
for $0<x<1$ and $|\Re \nu|<1$. Using Eq.~(\ref{eq:int}) one thus finds
\begin{eqnarray}
(V\psi )(x)&=& \frac{\pi}{\sin (\pi \beta)}
+\pi \cot (\pi \beta) \left[\frac{\beta}{x}-1\right]
\left(\frac{x}{1-x}\right)^\beta 
\label{eq:megav}
\\ 
& &+\frac{\pi}{\sin (\pi \beta^*)}
+\pi \cot (\pi \beta^*) \left[ \frac{\beta^*}{1-x}-1\right]
\left(\frac{1-x}{x}\right)^{\beta^*}
\nonumber
\end{eqnarray}
and hence for the expectation value of the Hamiltonian
\begin{equation}
\frac{ \langle \psi |H|\psi \rangle}{\langle\psi |\psi\rangle}
= 1 + \frac{m^2}{a} + \frac{a^2+b^2}{3a}\pi^2 - \theta_{eff}\frac{b}{a}
+{\cal O}(\beta^2)
\label{eq:exph}
\end{equation}
where $\beta = a+ib$ with $a$ and $b$ real.
This expression is minimized for
\begin{equation}
\beta_{min} = \frac{\sqrt{3}}{\pi} \sqrt{m^2 -
\frac{3 \theta_{eff}^2}{4\pi^2}}
+i \theta_{eff}\frac{3}{2\pi^2},
\label{eq:betamin}
\end{equation}
which agrees with Eq. (\ref{eq:bchiral}).
Substituting (\ref{eq:betamin}) into (\ref{eq:exph}) yields
for the invariant mass of the meson
\begin{equation}
M^2\equiv
\left.\frac{ \langle \psi |H|\psi \rangle}{\langle\psi |\psi \rangle}
\right|_{min}
= 1 +\frac{2\pi}{\sqrt{3}}
\sqrt{ m^2 - \frac{3}{4\pi^2}\theta_{eff}^2},
\label{eq:m2var}
\end{equation}
which is valid up to order ${\cal O}(m^2)$ and for 
$\theta_{eff} < m\frac{2\pi}{\sqrt{3}}$.
In order to turn Eq. (\ref{eq:m2var}) into a prediction for the $\theta$
dependence of the mass, we need to evaluate the schreening
factor (\ref{eq:thetaeff}). For this purpose, we note that
\begin{equation}
g_0^V =1 -{\cal O}(\beta^2),
\end{equation}
which, by completeness $\sum_n \left(g_n^V\right)^2=1$, also implies that
\begin{equation}
\sum_{n\neq 0} \left(g_n^V\right)^2 = {\cal O}(\beta^2).
\end{equation}
As a result, up to ${\cal O}(\beta^2)$, screening depends only on
the mass shift of the meson,
i.e.
\begin{equation}
\frac{\theta_{eff}}{\theta}=M^2-1 .
\end{equation}
To lowest order in $\theta$, one thus immediately obtains
\begin{equation}
\theta_{eff} = \theta \frac{2\pi m}{\sqrt{3}} +
{\cal O}(\theta^3),
\label{eq:t3}
\end{equation}
yielding 
\begin{equation}
M^2(\theta,m) = 1 + \frac{2\pi m}{\sqrt{3}}
\left(1-\frac{\theta^2}{2}\right) + {\cal O}(m^2) + 
{\cal O}(\theta^4),
\label{eq:m2a}
\end{equation}
which agrees to this order in $\theta$
with the result from bosonization \cite{adam}
\begin{equation}
M^2(\theta,m)_{bos} = 1 + 2e^\gamma m \cos (\theta)+ {\cal O}(m^2) 
\label{eq:t2bos}
\end{equation}
within 2 \% . 

One might suspect that the 2 \% difference between
Eq. (\ref{eq:m2a}) and Eq. (\ref{eq:t2bos}) is just a consequence
of using the ``wrong'' meson mass in Eq. (\ref{eq:thetaeff}).\
However, this is not the case, since the screening factor enters
Eq. (\ref{eq:m2a}) quadratically, and using $\mu_0^2$ from
bosonization instead of from the Bergknoff equation results in
a $\theta$-dependence in Eq. (\ref{eq:m2a}) which is 2 \% too
small instead of 2 \% too large. Nevertheless, we believe that
understanding the 2 \% puzzle in the $\theta=0$ case \cite{heinzl}
will eventually also help to understand the 2 \% deviation of
the $\theta$ dependent term above.

Much more important than the 2 \% deviation is the fact that above
calculation gave the correct result within 2 \% --- especially since
the calculation without screening corrections would have given a
$\theta$ dependence which diverges as $\theta \rightarrow 0$.
Only after inclusion of the screening factor did we obtain the
correct $m$-dependence, i.e. vanishing $\theta$-dependence in the
chiral limit. We consider this as a strong support for our
method of including screening effects into the effective background
field. Once more, we should emphasize that above result 
(\ref{eq:m2a}) was obtained
without including dynamical zero mode degrees of freedom.

\section{Summary}
We have studied the $\theta$ dependence of the meson mass in the
massive Schwinger model. The $\theta$ dependence is introduced via
a static background electric field. The essential new ingredient which
we introduce is a screened background field, which arises as an
effective coupling to fermion anti-fermion pairs with vanishing 
momentum. In the chiral limit we obtain results which are consistent
with results based on chiral perturbation theory.

While above results have been very reassuring, many open questions
remain, including deriving the effective Hamiltonian and calculating
meson masses for $\theta \approx \pi$. Such results would be particularly
interesting for investigations of ``vacuum periodicity'', which should
manifest itself in the effective Hamiltonian formalism as critical
behavior for $|\theta |= \pi$. Work in this direction is 
in progress\cite{us}.

\acknowledgements
M.~B. was supported in part by the D.O.E. (grant no. DE-FG03-96ER40965), 
by TJNAF and by an
internationalization supplementary grant from NMSU. 
K.~H. was partially supported by Sumitomo Foundation (No. 960517).

\end{document}